\documentclass[journal,twoside,a4paper]{IEEEtran}
\usepackage[left=1.5cm, right=1.5cm, top=1.5cm, bottom=1.5cm, includeheadfoot]{geometry}
\usepackage{multirow}
\usepackage{booktabs}
\usepackage{amsfonts}
\usepackage{placeins}
\usepackage{array}
\usepackage{amsmath,cases,amssymb}
\usepackage{amsmath}
\usepackage{graphicx}
\usepackage{cite}
\usepackage{subfigure}
\usepackage{epsfig}
\usepackage{url}
\usepackage{algorithm}
\usepackage{algorithmic}
\usepackage{epstopdf}
\usepackage{balance}
\usepackage{color}
\usepackage{algorithm}
\DeclareGraphicsExtensions{.eps,.pdf}

\newcommand{\st}{{\mathrm{s.t.}}}

\newcommand{\calK}{{\mathcal {K}}}
\newcommand{\non}{\nonumber}
\newcommand{\bw}{\mathbf {w}}

\usepackage{setspace}

\begin{document}
	\pagenumbering{gobble}
\title{Dynamic User Pairing for Non-Orthogonal Multiple Access in Downlink Networks}
\author{Kha-Hung Nguyen, Hieu V. Nguyen, Van-Phuc Bui, and Oh-Soon Shin*\\
	School of Electronic Engineering $\&$ Department of ICMC  Convergence Technology, Soongsil University, Korea \\ *Corresponding author (E-mail: osshin@ssu.ac.kr)
	\vspace{-0.9cm}
	\thanks{This work was supported in part by the Institute of Information $\&$ Communications Technology Planning $\&$ Evaluation (IITP) grant funded by the Korean government (MSIT) (No. 2017-0-00724, Development of Beyond 5G Mobile Communication Technologies (Ultra-Reliable, Low-Latency, and Massive Connectivity) and Combined Access Technologies for Cellular-based Industrial Automation Systems), and in part by the National Research Foundation of Korea (NRF) grant funded by the Korean government (MSIT) (No. 2019R1A2C1084834).
	}
	
}
\maketitle

\begin{abstract}
	This paper considers a downlink (DL) system where non-orthogonal multiple access (NOMA) beamforming and dynamic user pairing are jointly optimized to maximize the minimum throughput of all DL users. The resulting problem belongs to a class of mixed-integer non-convex optimization. To solve the problem, we first relax the binary variables to continuous ones, and then devise an iterative algorithm based on the inner approximation method which provides at least a local optimal solution. Numerical results verify that the proposed algorithm outperforms other ones, such as conventional beamforming, NOMA with random-pairing and heuristic-search strategies.
\end{abstract}
	
\section{Introduction}
Non-orthogonal multiple access (NOMA) has recently been considered as a promising technology for future mobile networks \cite{NOMA1, NOMA4,NOMAfor5G_beyond}. Unlike orthogonal multiple access (OMA), NOMA system allows multiple user equipments (UEs) to share the same time-frequency resource, thereby improving the system performance. There are two kinds of NOMA techniques: power-domain NOMA and code-domain NOMA, among which the former is more popularly utilized. Therefore, the term, NOMA, is often used to stand for the power-domain NOMA, and we follow the convention in this paper. From the key idea of NOMA, successive interference cancellation (SIC) technique is applied to a group of UEs so that the system performance is improved. For example, with a pair of two DL UEs, UE with better channel condition is equipped with an SIC receiver to decode and subtract out the message intended to the other UE before decoding its own message. This allows the base station (BS) to allocate more power to the user with worse channel condition.

Although many works have generalized the UE clustering to utilize the benefits of NOMA \cite{Hieu_fullduplex_NOMA_multizone, Ali:IEEEAccess:2016}, UE pairing scheme is more suitable and efficient in a particular scenario, i.e., a medium-range, due to the requirement of differences among the channel conditions of UEs \cite{PairingDLNOMA2zone}. A UE scheduling method with random pairing has been proposed in \cite{ChenCOMLL18_userpairing_scheduling}. In \cite{pairing1, NOMA_Subcarrier_PairFixIndex}, an optimal UE paring scheme has been investigated under the broadcast channel model with a single-antenna BS, where two unpaired UEs with the best and the worst channel gains are grouped together until all UEs are paired. The authors in \cite{NOMA_Subcarrier_PairFixIndex} have proposed a method in which two unpaired users with the best channel gains are paired successively. In \cite{Pairing2zonePhuc}, an optimal pairing scheme has been proposed in a two-zone cellular system, where one UE in the inner zone is paired with one UE in the outer zone. However, the UE pairing schemes in the previous strategies aim at assigning UE pairs to the different resource blocks \cite{pairing1, NOMA_Subcarrier_PairFixIndex} or clustering UEs in spatially-separated zones \cite{Pairing2zonePhuc}. Therefore, these schemes might not fully exploit the channel capacity as well as the advantages of NOMA, which motivate us to incorporate beamforming into UE pairing in NOMA systems.

In this paper, we consider a max-min rate optimization problem for a DL system, where a NOMA-based beamforming design is applied to the pairs of UEs without the requirement of different zones. Accordingly, any two UEs in the whole cell can be paired with each other, which will be referred to as dynamic global UE pairing (DGUP). This approach can reap the following advantages: \textit{(i)} enhancing the feasible region of UE pairs to improve the system performance; \textit{(ii)} enabling a hybrid beamforming design, with UEs paired for SIC or not. However, the distances from UEs to BS must be taken into account, such that the roles of two UEs in a pair are determined in priority. To manage the UE paring, we introduce binary variables which are restricted by a triangle assignment matrix. The resulting max-min rate problem belongs to a class of mixed-integer non-convex programming. To efficiently solve the problem, we first relax the binary variables into continuous ones. Then, we apply inner approximation (IA) method to derive a successive convex program \cite{IA}, which is solved by a low-complexity iterative algorithm to obtain at least a local optimal solution. Numerical results are provided to verify the effectiveness of the proposed algorithm in terms of both the performance and complexity.

\textit{Notation:} $\mathbf{h}^{H}$ is the Hermitian transpose of a vector $\mathbf{h}$. $\mathbb{C}$ and $\mathbb{R}$ are the sets of all complex and real numbers, respectively. $\|\cdot\|$ returns the Euclidean norm of a vector. $\mathbb{E}\{ \cdot \}$ stands for the expectation and  $\Re\{.\}$ denotes the real part of a complex number.

\section{System Model and Problem Formulation}
\subsection{System Model}

\begin{figure}[t]
	\centering
	\includegraphics[width=0.48\textwidth]{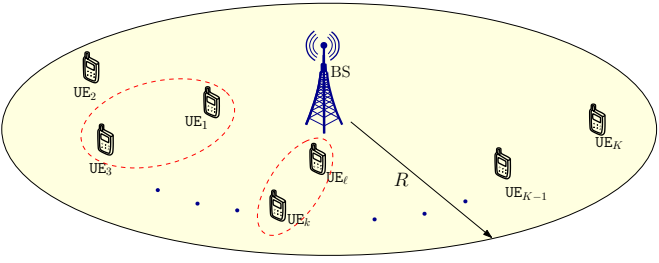}
	\caption{A DL NOMA system with $K$ DL users and one BS in a cell.}
	\label{fig:sysmodel}
\end{figure}

We consider a DL NOMA system where a BS equipped with $N$ antennas serves $K$ single-antenna UEs, with $\calK \triangleq \{1,2, \cdots, K\}$ and $\mathtt{UE}_{k}$ indicating the $k$-th UE. By using the principle of NOMA for a pair of UEs, the SIC is applied to the near UE, while the far UE still suffers from the interference caused by the near UE. To be convenient, we first sort UEs based on their distances to the BS in ascending order. Without loss of generality, we let $\mathtt{r}_{k}$ be the distance from the $k$-th UE to the BS after sorting. Then, the UEs' indices are re-numbered such that their distances satisfy the following non-decreasing condition as
\begin{IEEEeqnarray}{lll} \label{eq:ruser}
	\mathtt{r}_{k}\leq\mathtt{r}_{\ell},\quad \text{if } k<\ell, \forall k,\ell\in \calK.
\end{IEEEeqnarray}

The signal received at $\mathtt{UE}_{k}$ can be expressed as
\begin{IEEEeqnarray}{rCl}
	y_{k}=\sum_{k'\in \calK} \mathbf{h}_{k}^H \bw_{k'} {x}_{k'}+{n}_{k}, \quad {k} \in \calK, 
	\label{eq:yk}
\end{IEEEeqnarray}
where $\mathbf{h}_{k} \in \mathbb{C}^{N\times1}$ and $\bw_{k} \in \mathbb{C}^{N\times1}$ denote the vector of channel gain from the BS to $\mathtt{UE}_{k}$ and beamforming vector assigned to $ \mathtt{UE}_{k} $, respectively. We should note that the channel vectors involve both large- and small-scale fading. ${x}_{k} \text{ with }\mathbb{E}\{|{x}_{k}|^2\}=1$ denotes the symbol intended to $\mathtt{UE}_{k}$, while ${n}_{k} \sim \mathcal{CN}(0,\sigma_{k}^2)$ stands for the additive white Gaussian noise (AWGN) at $\mathtt{UE}_{k}$. For decoding the DL messages, we consider dynamic UE pairing to further improve the performance. Suppose that NOMA beamforming is applied to a pair of $\mathtt{UE}_{k}$ and $\mathtt{UE}_{\ell}$, $k,\ell \in \calK$. Let $\alpha_{k,\ell} \in \{0,1\},\forall k,\ell \in \calK$ indicate the user paring: 
\begin{equation}
\alpha_{k,\ell}=
\left 
\{
\begin{array}{ll}
1, &\quad \text{if } \mathtt{UE}_{k} \text{ and } \mathtt{UE}_{\ell} \text{ are paired and } \mathtt{UE}_{\ell} \text{'s} \\&\quad \text{message is decoded prior to }\mathtt{UE}_{k} \text{'s} \\&\quad \text{message at } \mathtt{UE}_{k} \\
0, &\quad \text{otherwise}.
\end{array}
\right.
\label{eq:pair_var}
\end{equation}
We define $\boldsymbol{\alpha} \triangleq [\alpha_{k,\ell}]_{k,\ell\in\calK}\in\{0,1\}^{K\times K}$. 
From \eqref{eq:ruser} and \eqref{eq:pair_var}, we intend to optimize an upper triangular matrix $\boldsymbol{\alpha}$:
\begin{IEEEeqnarray}{ll}
\begin{bmatrix}
\alpha_{1,1} & \alpha_{1,2} & \dots &  \alpha_{1,K}\\
\alpha_{2,1} & \alpha_{2,2} & \dots &  \alpha_{2,K}\\
\vdots		 & \vdots		& \ddots&  \vdots \\
\alpha_{K,1} & \alpha_{K,2} & \dots & \alpha_{K,K}\\
\end{bmatrix}
=
\begin{bmatrix}
0		&	\alpha_{1,2} 	& \dots  &\alpha_{1,K}\\
0		&	0			 	& \dots  &\alpha_{2,K}\\
\vdots 	& \vdots 	 		& \ddots &\vdots	  \\
0		& 	0			 	& \dots  & 	 		0\\
\end{bmatrix}.\quad
\end{IEEEeqnarray}
By letting $\bw \triangleq {[\bw_{k}]}_{k \in \calK}$, the signal-to-interference-plus-noise ratio (SINR) of $\mathtt{UE}_{k}$ is defined as	
\begin{IEEEeqnarray}{lll} \label{eq:SINR_k}
	\gamma_{k}(\bw, \boldsymbol{\alpha }) &=& \min {\left \{ \gamma_{0,k}(\bw,\boldsymbol{\alpha}),\underset{\ell\in\calK}{ \min} \{\gamma_{\ell,k}(\bw,\boldsymbol{\alpha})\} \right\}},
\end{IEEEeqnarray}
where $\gamma_{0,k}$ and $\gamma_{\ell,k}$ are SINR of $\mathtt{UE}_{k}$ at $\mathtt{UE}_{k}$ and at $\mathtt{UE}_{\ell}$, respectively, which are defined as
\begingroup
\allowdisplaybreaks
\begin{IEEEeqnarray}{lll}\label{eq:SINRkl}
	\gamma_{0,k}(\bw,\boldsymbol{\alpha}) &=& \frac{{ \left| \mathbf{h}_{k}^H \bw_{k} \right|}^2}{\Phi_{k}(\bw,\boldsymbol{\alpha})}, \IEEEyessubnumber \label{eq:SINRkl1} \\
	\gamma_{\ell,k}(\bw,\boldsymbol{\alpha}) &=& \frac{{ \left| \mathbf{h}_{\ell}^H \bw_{k} \right|}^2}{ \alpha_{\ell,k} \Psi_{\ell,k} (\bw)}, \IEEEyessubnumber \label{eq:SINRkl2}
\end{IEEEeqnarray}
\endgroup
where $\Phi_{k}(\bw,\boldsymbol{\alpha})$ and $\Psi_{k,\ell}(\bw)$ are defined as
\begin{IEEEeqnarray}{lll} \label{eq:IplusN}
	\Phi_{k}(\bw,\boldsymbol{\alpha}) &=& \sum_{\ell \in \calK\backslash \{k\}}{(1-\alpha_{k,\ell}){|\mathbf{h}_{k}^H \bw_{\ell}|}^2}+\sigma_{k}^2, \IEEEyessubnumber \label{eq:Phi} \\
	\Psi_{\ell,k}(\bw) &=& \sum_{\ell' \in \calK \backslash \{k\}} {|\mathbf{h}_{\ell}^H \bw_{\ell'}|^2 }+\sigma_{\ell}^2. \IEEEyessubnumber \label{eq:Si} 
\end{IEEEeqnarray}
Clearly, when $\alpha_{\ell,k}=0$, $\forall k, \ell\in \calK$, we obtain SINR of $\mathtt{UE}_{k}$ as  $\gamma_{k}(\bw,\boldsymbol{\alpha})=\gamma_{0,k}(\bw,\boldsymbol{\alpha})$.
\subsection{Problem Formulation}
The achievable rate at $\mathtt{UE}_{k}$ is given as (nats/s/Hz)
\begin{IEEEeqnarray}{lll}
	R_{k}(\bw,\boldsymbol \alpha) = \ln(1+\gamma_{k}(\bw,\boldsymbol{\alpha})), \quad k \in \calK. \label{eq:rate}
\end{IEEEeqnarray}
Therefore, the optimization problem for maximizing the minimum rate among all UEs (MMR for short) can be mathematically formulated as
\begin{IEEEeqnarray}{ll} \label{eq:maxminrate}
	\underset{\bw,\boldsymbol{\alpha}}{\max} \quad & \min_{k \in\calK} R_{k}(\bw,\boldsymbol{\alpha}) \IEEEyessubnumber \label{eq:maxzminratea}\\
\st& \underset{k\in\calK}{\sum}{\|\bw_{k}\|^2} \leq P^{\max}_{\text{BS}}, 									\quad \IEEEyessubnumber \label{eq:maxzminrateb}\\
&	\alpha_{k,\ell} \in \{0,1\},														\quad \IEEEyessubnumber \label{eq:maxzminratec}\\
&	\sum_{k \in\calK}{\alpha_{k,\ell}} \leq 1,\quad \forall \ell \in \calK, 	\quad \IEEEyessubnumber \label{eq:maxzminrated}\\
&	\sum_{\ell \in\calK}{\alpha_{k,\ell}} \leq 1,\quad \forall k \in \calK, 	\quad \IEEEyessubnumber \label{eq:maxzminratee}\\
&	\alpha_{k,k}=0, \quad \forall k \in \calK , 							\quad \IEEEyessubnumber \label{eq:maxzminratef}\\
&	\alpha_{k,\ell}=0, \quad \forall k,\ell \in\calK\mid k > \ell, 					\quad \IEEEyessubnumber \label{eq:maxzminrateg}\\
&	\alpha_{k,\ell}+\underset{k'\in\calK}{\sum}{\alpha_{\ell,k'}} \leq 1,\quad \forall k,\ell \in\calK\mid k < \ell, \quad \IEEEyessubnumber \label{eq:maxzminrateh}\\
&	\alpha_{k,\ell}+\underset{\ell'\in\calK}{\sum}{\alpha_{\ell',k}} \leq 1,\quad \forall k,\ell \in\calK\mid k < \ell, \quad \IEEEyessubnumber \label{eq:maxzminratei}
\end{IEEEeqnarray}

Constraint \eqref{eq:maxzminrateb} ensures that the total transmit power at the BS does not exceed maximum power budget, $P^{\max}_{\text{BS}}$. Constrains \eqref{eq:maxzminratec}-\eqref{eq:maxzminratei} are the criteria for UE pairing. Constraints \eqref{eq:maxzminrated}, \eqref{eq:maxzminratee}, \eqref{eq:maxzminrateh} and \eqref{eq:maxzminratei} guarantee that each UE is paired with at most one other UE, while constrains \eqref{eq:maxzminratef} and \eqref{eq:maxzminrateg} assign zeros to the lower triangular entities of $ \boldsymbol{\alpha } $. However, the objective function \eqref{eq:maxzminratea} is non-convex and \eqref{eq:maxzminratec}-\eqref{eq:maxzminratei} are the binary constraints. Therefore, problem \eqref{eq:maxminrate} belongs to a mixed-integer non-convex problem. We should emphasize that the use of $\boldsymbol{\alpha }$ allows arbitrary two users in the cell to be paired. We refer to this strategy as \textit{dynamic user pairing}.

%
\section{Proposed Iterative Algorithm}
To efficiently solve problem \eqref{eq:maxminrate}, we first relax binary variable $\alpha_{k,\ell} \in \{0,1\}$ to $0 \leq \alpha_{k,\ell} \leq 1$ $\forall k,\ell \in \calK$. 
Then, problem \eqref{eq:maxminrate} can be rewritten as the following tractable form:
\begin{IEEEeqnarray}{lll} \label{eq:maxminrate2}
	\underset{\bw,\boldsymbol{\alpha},\eta}{\max} \quad & \eta 	\quad \IEEEyessubnumber \label{eq:maxminrate2a} \\
\st & R_{k} \geq \eta, \quad \forall k \in\calK, \quad \IEEEyessubnumber \label{eq:maxminrate2b}\\
& 	0 \leq \alpha_{k,\ell} \leq 1, \quad \forall k,\ell \in \calK, \quad \IEEEyessubnumber \label{eq:maxminrate2c}\\
&	\eqref{eq:maxzminrateb},\eqref{eq:maxzminrated}-\eqref{eq:maxzminratei}. \quad \IEEEyessubnumber  \label{eq:maxminrate2d}
\end{IEEEeqnarray}
However, constraint \eqref{eq:maxminrate2b} is still non-convex. Therefore, we aim to handle this constraint as follows.

\textit{\underline{Convexifying constraint \eqref{eq:maxminrate2b}:}} First, constraint \eqref{eq:maxminrate2b} can be written as
\begin{subnumcases} {\label{eq:ln_Rk_geq_eta}}
	 \ln(1+\gamma_{0,k}(\bw,\boldsymbol{\alpha}))\geq \eta , \quad k \in \calK, \IEEEyessubnumber \label{eq:ln_Rk_geq_etaa}\\
	 \ln(1+\gamma_{\ell,k}(\bw,\boldsymbol{\alpha}))\geq \eta , \quad k \in \calK, \IEEEyessubnumber \label{eq:ln_Rk_geq_etab}
\end{subnumcases}
By utilizing \cite[Eq. (20)]{Dinh_TCOM_17}, a lower bound of the left-hand side of \eqref{eq:ln_Rk_geq_etaa} at the iteration $(i+1)$ around a feasible point $(\bw^{(i)},\boldsymbol{\alpha}^{(i)})$ is given by
\begin{IEEEeqnarray}{ll}
	\ln(1+\gamma_{0,k}(\bw,\boldsymbol{\alpha})) \geq  f_{0,k}^{(i)} + 2f_{1,k}^{(i)}(\bw) -f_{2,k}^{(i)}(\bw,\boldsymbol{\alpha}), \label{eq:boundR_0k}
\end{IEEEeqnarray}
where $f_{0,k}^{(i)}$, $f_{1,k}(\bw^{(i)},\boldsymbol{\alpha}^{(i)})$ and $f_{2,k}^{(i)}(\bw,\boldsymbol{\alpha})$ are respectively defined as
\begin{IEEEeqnarray}{lll}
	f_{0,k}^{(i)}&\triangleq&\ln(1+\gamma_{0,k}(\bw^{(i)},\boldsymbol{\alpha}^{(i)}))-\gamma_{0,k}(\bw^{(i)},\boldsymbol{\alpha}^{(i)}), \nonumber \\
	f_{1,k}^{(i)}(\bw)&\triangleq&\frac{\mathfrak{R}\{(\mathbf{h}_{k}^H \bw_{k}^{(i)})*(\mathbf{h}_{k}^H \bw_{k})\}}{\Phi_{k}(\bw^{(i)},\boldsymbol{\alpha}^{(i)})}, \nonumber \\
	f_{2,k}^{(i)}(\bw,\boldsymbol{\alpha})&\triangleq&|\mathbf{h}_{k}^H \bw_{k}|^2 \Xi_{k}^{(i)}  +\sigma_{k}^2 \Xi_{k}^{(i)}	\nonumber\\
	&&+\underset{\ell\in \calK \backslash\{k\}}{\sum}{(1-\alpha_{k,\ell})|\mathbf{h}_{k}^H \bw_{\ell}|^2 \Xi_{k}^{(i)}},		\nonumber
\end{IEEEeqnarray}
with $\Xi_{k}^{(i)}$ is defined as
\begin{IEEEeqnarray}{ll}
	\Xi_{k}^{(i)}\triangleq\Phi_{k}(\bw^{(i)},\boldsymbol{\alpha}^{(i)})^{-1} -(\Phi_{k}(\bw^{(i)},\boldsymbol{\alpha}^{(i)})+|\mathbf{h}_{k}^H \bw_{k}^{(i)}|^2)^{-1}. \nonumber
\end{IEEEeqnarray}
It can be seen that $f_{2,k}^{(i)}$ is still non-convex, leading to a non-convex constraint \eqref{eq:boundR_0k}. To approximate \eqref{eq:boundR_0k}, we introduce new variable $\boldsymbol{\mu} \triangleq \{\mu_{k,\ell}\}_{k,\ell\in \calK}$, which satisfies the following convex constraint:
\begin{equation} \label{eq:bound_mu_k_ell}
	|\mathbf{h}_{k}^H \bw_{\ell}|^2 \leq \mu_{k,\ell}, \ \forall k, \ell \in \calK.
\end{equation}
Consider a function $ v(x,z)=xz $ with $ x>0,z>0 $. By using \cite[Eq. (B.1)]{DinhJSAC18}, an upper bound of $ v(x,y) $ is given as
\begin{IEEEeqnarray}{ll}
	v(x,z) \leq \frac{x^{(i)}}{2 z^{(i)}}z^2+\frac{z^{(i)}}{2x^{(i)}}x^2 \triangleq \hat{v}(x,z).		\nonumber
\end{IEEEeqnarray}
Then, the upper bound of $(1-\alpha_{k,\ell}) \mu_{k,\ell}$ can be expressed as
\begin{IEEEeqnarray}{ll}\label{eq:bound_1_alpha_mu}
	(1-\alpha_{k,\ell}) \mu_{k,\ell} &\leq \hat{v}(1-\alpha_{k,\ell},\mu_{k,\ell}). 
\end{IEEEeqnarray}
From \eqref{eq:bound_mu_k_ell} and \eqref{eq:bound_1_alpha_mu}, an upper bound of $f_{2,k}^{(i)}(\bw,\boldsymbol{\alpha})$ is given by
\begin{IEEEeqnarray}{lll}
	f_{2,k}^{(i)}(\bw,\boldsymbol{\alpha}) &\leq& |\mathbf{h}_{k}^H \bw_{k}|^2 \Xi_{k}^{(i)} +\sigma_{k}^2 \Xi_{k}^{(i)} \nonumber \\ &&
				+\underset{\ell\in \calK \backslash\{k\}}{\sum}{(1-\alpha_{k,\ell}) \mu_{k,\ell} \Xi_{k}^{(i)}} \nonumber\\
	&\leq& |\mathbf{h}_{k}^H \bw_{k}|^2 \Xi_{k}^{(i)}+\sigma_{k}^2 \Xi_{k}^{(i)} \nonumber \\
			&&+\underset{\ell\in \calK \backslash\{k\}}{\sum}{ \hat{v}(1-\alpha_{k,\ell},\mu_{k,\ell}) \Xi_{k}^{(i)}} \nonumber \\
	&\triangleq& \hat{f}_{2,k}^{(i)}(\bw,\boldsymbol{\alpha},\boldsymbol{\mu}). \label{eq:boundf2k}
\end{IEEEeqnarray}
Finally, constraint \eqref{eq:ln_Rk_geq_etaa} is convexified as
\begin{IEEEeqnarray}{ll}
	f_{0,k}^{(i)} + 2f_{1,k}^{(i)}(\bw) -\hat{f}_{2,k}^{(i)}(\bw,\boldsymbol{\alpha},\boldsymbol{\mu}) \geq \eta, \ \forall k, \ell \in \calK. \label{eq:R0k_geq_eta}
\end{IEEEeqnarray}
Similarly, constraint \eqref{eq:ln_Rk_geq_etab} is approximated as
\begin{IEEEeqnarray}{ll}
	g_{0,\ell,k}^{(i)} + 2g_{1,\ell,k}^{(i)}(\bw) -\hat{g}_{2,\ell,k}^{(i)}(\bw,\boldsymbol{\alpha},\boldsymbol{\mu}) \geq \eta, \ \forall k, \ell \in \calK, \label{eq:Rlk_geq_eta}
\end{IEEEeqnarray}
where $ g_{0,\ell,k}^{(i)} $, $g_{1,\ell,k}^{(i)}(\bw)$ and $\hat{g}_{2,\ell,k}^{(i)}(\bw,\boldsymbol{\alpha},\boldsymbol{\mu})$ are respectively given as

\begingroup
\allowdisplaybreaks
\begin{IEEEeqnarray}{lll}
	g_{0,\ell,k}^{(i)}&\triangleq&\ln(1+\gamma_{\ell,k}(\bw^{(i)},\boldsymbol{\alpha}^{(i)}))-\gamma_{\ell,k}(\bw^{(i)},\boldsymbol{\alpha}^{(i)}), \nonumber \\
	g_{1,\ell,k}^{(i)}(\bw)&\triangleq&\frac{\mathfrak{R}\{(\mathbf{h}_{\ell}^H \bw_{k}^{(i)})*(\mathbf{h}_{\ell}^H \bw_{k})\}}{\alpha_{\ell,k}^{(i)}\Psi_{\ell,k}(\bw^{(i)})}, \nonumber \\
	\hat{g}_{2,\ell,k}^{(i)}(\bw,\boldsymbol{\alpha},\boldsymbol{\mu})	&\triangleq&|\mathbf{h}_{\ell}^H \bw_{k}|^2 \Theta_{\ell,k}^{(i)} + \alpha_{\ell,k} \sigma_{\ell}^2 \Theta_{\ell,k}^{(i)} \nonumber \\
			&&+\underset{\ell'\in \calK \backslash\{k\}}{\sum}{\hat{v}(\alpha_{\ell,k},\mu_{\ell,\ell'}) \Theta_{\ell,k}^{(i)}}, \nonumber 
\end{IEEEeqnarray}\endgroup
with $ \Theta_{\ell,k}^{(i)} $ is defined as
\begin{IEEEeqnarray}{lll}
	\Theta_{\ell,k}^{(i)}\triangleq(\alpha_{\ell,k}^{(i)}\Psi_{\ell,k}(\bw^{(i)}))^{-1} -(\alpha_{\ell,k}^{(i)}\Psi_{\ell,k}(\bw^{(i)})+|\mathbf{h}_{\ell}^H \bw_{k}^{(i)}|^2)^{-1}. \non
\end{IEEEeqnarray}

In summary, successive convex program at iteration $(i+1)$ to solve problem \eqref{eq:maxminrate} is formulated as
\begin{IEEEeqnarray}{rl} \label{eq:final_problem}
	\underset{\bw,\boldsymbol{\alpha},\boldsymbol{\mu},\eta}{\max} \quad& \eta	\IEEEyessubnumber \\
\st\quad 
&	\eqref{eq:maxzminrateb},\eqref{eq:maxzminrated}-\eqref{eq:maxzminratei},\eqref{eq:maxminrate2c},\eqref{eq:bound_mu_k_ell}, \eqref{eq:R0k_geq_eta},\eqref{eq:Rlk_geq_eta}.		\IEEEyessubnumber
\end{IEEEeqnarray}

It can be foreseen that the values of $\boldsymbol{\alpha}$ may not be binary at the convergence due to the relaxation, leading to a violation of the constraint \eqref{eq:maxzminratec}. Therefore, we use a rounding function to recover the binary values as
\begin{IEEEeqnarray}{ll} \label{eq:roudingalpha}
	\alpha_{k,\ell}^\star=\lfloor \alpha_{k,\ell}^{(i)} + \frac{1}{2} \rfloor,\quad \forall k,\ell \in\calK.
\end{IEEEeqnarray}
Summarily, the proposed iterative algorithm is showed in Algorithm 1. Since Algorithm 1 is executed using IA method, it converges at a stationary point, which satisfies the Karush-Kuhn-Tucker (KKT) invexity.

\begin{algorithm}[t]
	\begin{algorithmic}[1]
		\protect\caption{Proposed Algorithm to Solve Problem \eqref{eq:maxminrate2}}
		\label{alg_1}
		\STATE{\textbf{Initialization:}} Set $i=0$ and randomly generate an initial feasible point $(\mathbf{w}^{(0)}, \boldsymbol{\alpha}^{(0)}, \boldsymbol{\mu}^{(0)})$.\\
		\REPEAT
		\STATE Solve \eqref{eq:final_problem} to obtain the optimal point $(\mathbf{w}^\star,\boldsymbol{\alpha}^\star, \boldsymbol{\mu}^\star)$.
		\STATE Update\ \ $  (\mathbf{w}^{(i+1)},\boldsymbol{\alpha}^{(i+1)},\boldsymbol{\mu}^{(i+1)}) := (\mathbf{w}^\star,\boldsymbol{\alpha}^\star,\boldsymbol{\mu}^\star)$.
		\STATE Set $i=i+1.$
		\UNTIL Convergence\\
		\STATE Update $\boldsymbol{\alpha}^\star$ by using rounding function \eqref{eq:roudingalpha}.\\
		\STATE Reset $ i $ and repeat steps 2-6 to get optimal beamforming vector $\mathbf{w}^\star$ corresponding to the given binary variables $\boldsymbol{\alpha}^\star$.\\
		\STATE \textbf{Output:} ($\mathbf{w}^\star, \boldsymbol{\alpha}^\star$).
\end{algorithmic} \end{algorithm}

\textit{Complexity analysis}: The complexity of solving \eqref{eq:final_problem} is $\mathcal{O}\big(x^{2.5}(y^2 + x)\big)$ \cite{sedumi}, where $x=6K^2+3K+1$ denotes the number of quadratic/linear constraints and $y=NK+K^2+1$ denotes that of variables.

\section{Numerical Results}
To evaluate the proposed algorithm, we consider a system consisting of one BS and $ K=8 $ UEs. The UEs are randomly distributed in a small-cell with radius $ R=200 $ m. Other simulation parameters are provided in Table I.
Algorithm 1 is terminated when the difference of the objective values in two consecutive iterations becomes smaller than $ 10^{-3} $.
The proposed scheme (Alg. 1) is compared with four existing UE pairing schemes for NOMA, where the values of $ \boldsymbol{\alpha } $ are determined as follows.
\begin{itemize}
	\item \textit{``Random Pairing''}: After randomly setting the upper triangular matrix $ \boldsymbol{\alpha } $, Algorithm 1 is used to handle the power control.
	\item \textit{``Scheme 1''}: As in \cite{NOMA_Subcarrier_PairFixIndex}, UE $2k$ is paired with UE $(2k-1)$, i,e., $\alpha_{2k-1,2k}=1, 1\leq k\leq \lfloor 0.5K \rfloor$.
	\item \textit{``Scheme 2''}: As in \cite{pairing1}, UE $k$ is paired with UE $(K-k+1)$, i,e., $\alpha_{k,K-k+1}=1$, $1\leq k\leq \lfloor 0.5K \rfloor$.
	\item \textit{``Greedy Pairing''}: Greedy strategy aims to maximize the number of UE pairs as in the two-zone model. Specifically, UE $ k $ is paired with UE ($ K- \lfloor 0.5K \rfloor +k$),  $ 1\leq k \leq \lfloor0.5K\rfloor $.
\end{itemize}
After fixing $ \boldsymbol{\alpha } $, Algorithm 1 is applied to maximize the minimum rate of all UEs under the power control. Furthermore, we examine two baseline schemes: \textit{(i)} \textit{``Exhaustive Search''} where all possible values of $ \boldsymbol{\alpha } $ are considered to derive the subproblems. Then, we apply Algorithm 1 with the fixed values of $ \boldsymbol{\alpha } $  to solve the subproblems before selecting the best solution; \textit{(ii)} \textit{``Beamforming''} where $ \boldsymbol{\alpha} $ is set to zero, and thus, Algorithm 1 is utilized for conventional beamforming design.

\begin{table}[t]
	\caption{Simulation Parameters}
	\label{parameter}
	\centering
		\scalebox{0.8}{
		\begin{tabular}{l|l}
			\hline\hline
			Parameter & Value \\
			\hline
			System bandwidth                             &  20 MHz \\
			Noise power spectral density & -174 dBm/Hz \\
			Path loss between BS and UEs, $\sigma_{\mathsf{PL}}$   & 145.4 + 37.5$\log_{10}(d)$ dB\\
			Radius of the cell $(R)$ &  200 m\\
			Distance limit from BS to the nearest UE $ (d) $ & $\geq$ 10 m\\
			\hline		   				
	\end{tabular}}
\end{table}	

\begin{figure}
	\centering
	\includegraphics[width=0.84\columnwidth]{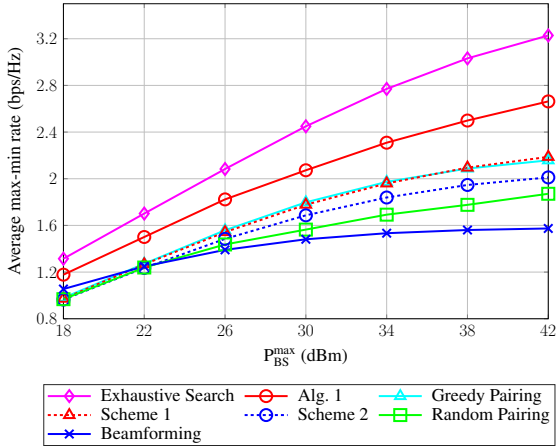}
	\vspace{-0.5pt}
	\caption{Average max-min rate performance versus $ P_{\mathrm{BS}}^{\max} $ with $ N=4 $.}
	\label{fig:MMR_Pbs}
\end{figure}

\begin{figure}
	\centering
	\includegraphics[width=0.84\columnwidth]{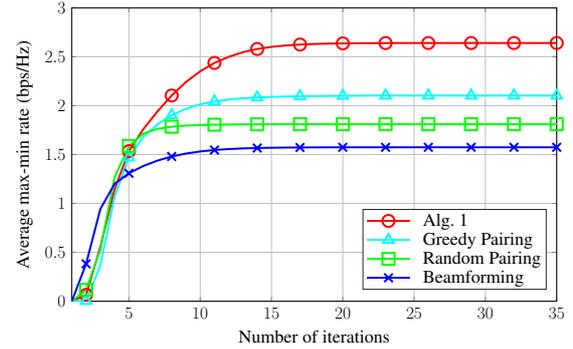}
	\vspace{-5pt}
	\caption{Convergence behavior of Alg. 1 for one channel realization with $ P_{\mathrm{BS}}^{\max}=38 \text{dBm}$ and $ N=4 $.}
	\label{fig:convergence}
\end{figure}

The MMR performance of the proposed scheme and the other six schemes mentioned above are shown in Fig.~\ref{fig:MMR_Pbs}. Compared to other strategies except for exhaustive search, Algorithm 1 has higher MMR; The gains against greedy pairing, random pairing and beamforming are about $ 0.5 $bps/Hz, $ 0.8 $bps/Hz and $ 1.1 $bps/Hz, respectively. This demonstrates the effectiveness of a joint optimization of dynamic user pairing and beamforming. Fig.~\ref{fig:convergence} depicts the convergence behaviors of the proposed algorithm and other strategies. It can be observed that all schemes attain 90\% performance after 15 iterations, which verifies the effectiveness of the proposed method in terms of complexity. 



\section{Conclusion}
We have studied a joint optimization of NOMA beamforming and dynamic UE pairing. The formulated max-min rate problem is a mixed-integer non-convex program. To efficiently solve the problem, we derive an iterative algorithm based on the relaxation and IA methods to obtain at least a local optimal solution. Numerical results demonstrate that the proposed algorithm outperforms the existing methods.

\setstretch{0.98}
\bibliographystyle{IEEEtran}
\bibliography{Journal}
\end{document}